# Gyrokinetic simulations of the effects of magnetic islands on microturbulence in KSTAR

Xishuo Wei[1*], Javier H Nicolau[1], Gyungjin Choi[1,2], Zhihong Lin[1], Seong-Moo Yang[3], SangKyeun Kim[3,4], WooChang Lee[5], Chen Zhao[6], Tyler Cote[6], JongKyu Park[3,2], Dmitri Orlov[7]
1. University of California, Irvine, CA, United States of America
2. KAIST, Daejeon, Republic of Korea
3. Princeton Plasma Physics Laboratory, Princeton, NJ, United States of America
4. Princeton University, Princeton, NJ, United States of America
5. Korea Institute of Fusion Energy, Daejeon, Republic of Korea
6. General Atomics, San Diego, CA, United States of America
7. University of California, San Diego, CA, United States of America

*Email: xishuow@uci.edu

## Abstract

Gyrokinetic simulations are utilized to study effects of magnetic islands on the ion temperature gradient (ITG) turbulence in the KSTAR tokamak with resonant magnetic perturbations. Simulations show that the transport is controlled by the nonlinear interactions between the ITG turbulence and self-generated vortex flows and zonal flows, leading to an anisotropic structure of fluctuation and transport on the poloidal plane and in the toroidal direction. Magnetic islands greatly enhance turbulent transport of both particle and heat. The turbulent transport exhibits variations in the toroidal direction, with transport through the resonant layer near the island X-point being enhanced when the X-point is located at the outer mid-plane. A quantitative agreement is shown between simulations and KSTAR experiments in terms of time frequency and perpendicular wavevector spectrum.

## 1 Introduction

During the H-mode operation of tokamaks, type-I Edge Localized Modes (ELMs) generate transient energy bursts at the plasma boundary, which can degrade plasma confinement and cause erosion of plasma-facing materials. One promising technique to avoid or mitigate these instabilities is the application of Resonant Magnetic Perturbations (RMPs)[1], originally developed to control magnetohydrodynamic (MHD) instabilities near the resonant magnetic surfaces. By applying current to external coils surrounding the tokamak, magnetic perturbations with helicity matching the field lines at resonant surfaces are generated. The RMP control of ELM has been demonstrated on several tokamak devices, including DIII-D[2], JET[3], KSTAR[4], and EAST[5], and is also predicted to be effective in future devices like ITER[6]. ELM suppression via RMPs enables long-pulse, steady-state operations while maintaining high confinement parameters. For instance, a KSTAR discharge[7] with a normalized plasma kinetic pressure $\beta_N \approx 3$ was sustained for 12 seconds with an effective fusion gain $G > 0.4$ using an edge safety factor $q_{95} = 4$. However, RMPs can also cause undesirable side effects on turbulent transport, such as a reduction in plasma density at both the edge and core (density pump-out) and an increase in the power threshold required to access high confinement mode (H-mode)[4], [8]. Understanding the physics mechanism of these RMP effects on turbulent transport is crucial for optimizing RMP configurations to achieve better confinement and higher fusion gain during steady-state operations. A key aspect of this is understanding the impact of magnetic islands (MIs) induced by the RMPs on the turbulent transport.

The MIs can form through magnetic reconnection on the rational surfaces where rotational transform of the unperturbed magnetic fields matches the RMP helicity. Even small MIs can significantly alter flux surface topology and plasma transport[9]. The impacts of MIs on plasma behavior have been observed across various tokamak and stellarator devices[10], [11], [12], [13]. The reduction in the pressure gradient within the island region has been observed, accompanied by a decrease in the turbulence intensity. Strong electron temperature $T_e$ fluctuations have been observed near the X-point of the islands while the fluctuations near the island O-point are reduced. Turbulence can spread into the island region via the X-point. Some experiments have shown a potential positive role of MIs in the formation of internal transport barriers (ITBs)[10], [12]. Recent experiments have directly revealed the magnetic island effect on plasma flow, fluctuations and transport. A KSTAR experiment has shown that the inhomogeneous turbulence surrounding the MIs is related to variations in the ExB flow shear near the O-point and X-point[14]. In an HL-2A experiment, the modulation of turbulent fluctuation of electron temperature and density within the island region has been found[15], and a minimal island width has been identified for the modulation. In another HL-2A experiment, different flow profiles around O-point and X-point have been found[16], and especially, a large flow shear has been found at the outer separatrix of island near O-point. The causality of flow shear and fluctuation modulation has also been demonstrated. Similar spatial structure of poloidal flow and the effect of island on flow profile have been found in the W7-X stellarator experiment for the m=5,n=5 island.[17] Remarkably, a transport barrier has been demonstrated directly due to the m=5,n=5 island inside the last closed flux surface[18]. In addition, a TJ-II stellarator experiment[19] has also found the flow shear along the island separatrix, and its effect on fluctuation reduction and transport regulation, and the asymmetry of flow shear at the inner and outer sides of separatrix have been identified. Theoretical studies have also explored various aspects of MI effects on neoclassical transport[20] and bootstrap current[21], vortex flow generation[22], [23], and the long-term evolution of MIs[24]. Additionally, resonant interactions between particle motion and MIs have been studied theoretically [25], [26]. Despite these advances, the multiscale interactions between MIs and turbulent transport remain unsolved theoretically. There is no first-principles theory that can self-consistently predict turbulent transport in the presence of MIs. Numerical simulations are needed to further investigate the detailed mechanisms responsible for confinement degradation and to elucidate the spatial characteristics of turbulent transport.

Extensive numerical studies have been conducted to understand the effect of MI and 3D fields on turbulent transport. It has been demonstrated that the MIs can stabilize[27] and localize[28] the ion temperature gradient (ITG) modes. The generation of vortex flows by MIs in neoclassical tearing mode (NTM) simulations is examined in [29], which was complemented by the study of mechanism of turbulence suppression via vortex flows[30]. Additional research into mean flow generation by MIs has also been conducted[31], [32]. Several studies have also shown that MIs suppress turbulent transport within the island region[33], [34], identifying a critical island width necessary for transport reduction. This finding has been further corroborated by later gyrokinetic simulations, which also identified the role of island width in transport suppression[35]. The mechanism of turbulence spreading into the island region has been explored in [36] and [37]. More recently, the effects of neoclassical toroidal viscosity (NTV) have been examined, with a focus on density pump-out, and these results have been compared directly to experimental data.[38], [39] More interesting studies on the multi-scale interaction between evolving MIs and turbulence were highlighted in review papers[40], [41].

In addition to the aforementioned studies, the Gyrokinetic Toroidal Code (GTC)[42] has been extensively employed to investigate the effects of 3D magnetic fields from RMPs and the interaction between magnetic islands (MIs) and other modes. GTC has been verified for a wide range of physical phenomena across various devices, including tokamaks, stellarators[43], and field-reversed configurations (FRCs)[44], encompassing neoclassical transport [16], microturbulence[45], meso-scale Alfvén eigenmodes (AEs)[46], and macro-scale MHD modes[47]. The impact of MIs on profile flattening and linear ITG instability has been studied in [48]. The simulation model for tearing modes (TMs) was developed and TM instability in both the resistive and collisionless limits was verified[49], [50]. Further studies examined the effects of MIs on bootstrap current due to profile flattening within the island region[21]. The influence of electron cyclotron current drive on the evolution of MIs was investigated[51], [52]. The effects of MIs on ITG turbulent transport and the interactions between turbulence and MIs were studied in [53], [54]. The drift-tearing modes[55], [56] and double tearing modes[57] have been simulated. The interaction between energetic particles (EPs) and islands induced by NTMs was explored in [58]. More recently, the effect of island width on zonal flow generation and turbulent transport in a simple model tokamak was analyzed[59]. The NTM simulation was verified with the modified Rutherford equation and the influence of finite Larmor radius effects on NTM instability was studied[60]. In addition to these studies, GTC has also been employed to investigate the effects of non-resonant RMP fields on turbulent transport and the generation of radial electric fields[61], [62], [63]. These simulations offer valuable insights into the complex interplay between MIs, turbulence, and RMP fields, advancing our understanding of plasma transport in fusion devices.

In this work, we use the GTC simulation to study the effects of the MIs on the turbulent transport using a realistic geometry and profiles in an actual experiment, KSTAR long-pulse discharge #19118, which features a large magnetic island with a width of approximately 4 cm [64], [65] generated by the RMP that suppresses the ELMs. We performed global, self-consistent GTC electrostatic simulations using gyrokinetic ions and drift kinetic electrons. Our simulations verified the presence of anisotropic turbulence structures and turbulence spreading in the poloidal plane, regulated by nonlinear interactions between turbulence, zonal flows, and vortex flows. Magnetic islands greatly enhance turbulent transport of both particle and heat. We also observed that turbulent transport exhibits variations in the toroidal direction, with transport through the resonant layer near the X-point being enhanced when the X-point is located at the outer mid-plane. Furthermore, we compared the features of turbulence between GTC simulations and experimental observations, finding good agreement regarding the magnetic island effect on time frequency and perpendicular wavevector spectrum. The variation of density fluctuation at O-point and X-point are also similar to previous DIII-D[66] and TEXT experiments[67]. This work enhances our understanding of turbulent transport in the presence of MIs and provides insights in improving the confinement in RMP experiments.

This paper is organized as follows: Section 2 describes the KSTAR experimental data, the gyrokinetic simulation model including magnetic islands, and the simulation settings. Section 3 presents the results of the turbulence simulations. Finally, a summary is provided in Section 4.

## 2 Simulation model and simulation settings

The experimental data of KSTAR discharge #19118 at t=2950ms is presented in Fig.1. The electron temperature is measured through electron cyclotron emission and the electron density is measured by Thomson scattering. The ion temperature is measured by charge exchange spectroscopy. Then the smoothed profile is obtained by fitting the discrete data points. The n=1 RMP was applied at t=2400ms in the discharge, leading to the formation of the MI and the subsequent mode locking was observed at t=2800ms. At t=2950ms, the radial electron temperature profile exhibits a clear flattening effect near q=2 surface. The q=2 surface is located at $\rho_t = 0.549$, and the island region covers from $\rho_t = 0.42$ to $\rho_t = 0.65$ in the neighbor of O-points. Here, $\rho_t \equiv \sqrt{\psi_N}$ and $\psi_N$ is the normalized toroidal magnetic flux. Consistently, the Electron Cyclotron Emission Imaging (ECEI) data reveals the (m,n)=(2,1) MI structure at the q=2 surface. On the other hand, no clear flattening effect is observed in the electron density and ion temperature, possibly due to the large error in the measurements of these profiles. In the simulations, we consider a single ion species (Deuterium) and assume the equilibrium ion density $n_{i0}$ to be identical to electron density $n_{e0}$.

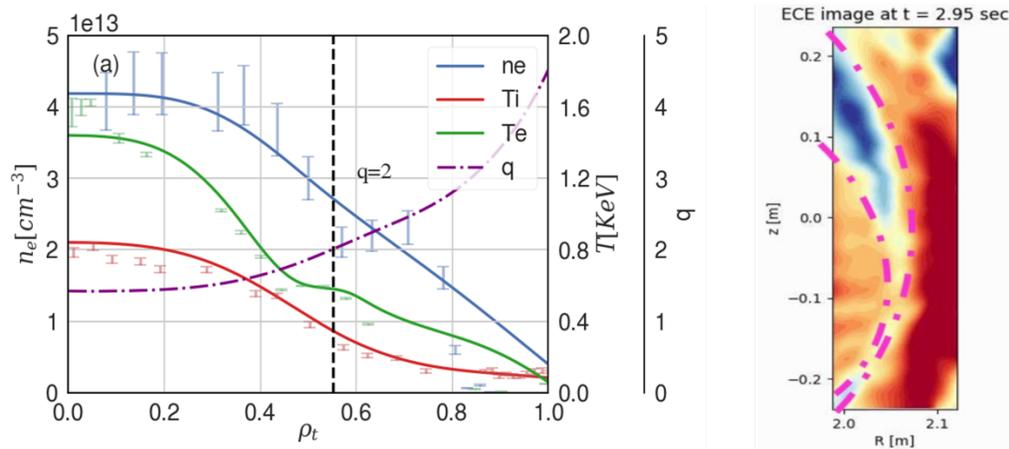

*Figure 1 (a) The radial profiles of electron temperature, electron density, ion temperature, and safety factor of KSTAR shot #19118 at t=2.95s. The vertical error bars stand for the measurement error. The RMP has been turned on and the MI has been formed. $n_i$*

*is assumed to be identical to $n_e$. (b) The temperature fluctuations on the R-Z poloidal plane measured by ECE imaging at t=2.95s. The two magenta curves show the separatrix of MI.*

GTC employs gyrokinetic equations for both ions and electrons to simulate the low-frequency waves and the associated turbulent transport. The static magnetic field can be expressed as $\mathbf{B} = \mathbf{B}_0 + \nabla \times (\alpha_{IS}\mathbf{B}_0)$, where $\mathbf{B}_0$ is the ambient magnetic field, and $\delta \mathbf{B}_{IS} \equiv \nabla \times (\alpha_{IS}\mathbf{B}_0)$ is the magnetic field induced by the static MI. Both of $\mathbf{B}_0$ and $\alpha_{IS}$ in the Boozer coordinate system are solved by M3D-C1 code[68] based on the KSTAR experimental data. The stationary state of $\mathbf{B}_0$ and $\alpha_{IS}$ is obtained by the M3D-C1 simulation and taken as the static background equilibrium for the GTC simulation. The shape of the island field is shown in Fig.2. The left panel illustrates that the maximum amplitude of $\alpha_{IS}$ is located at the X-points and the O-points. The right panel presents the Poincare plot electron guiding center orbit at the $\zeta = \pi$ plane, with the X-point situated at the outer mid-plane. The electron motion is confined to the constant perturbed flux $\psi_h$ surfaces, clearly displaying the (2,1) island structure since the electron guiding center orbit width is very small compared with the MI size. The helical flux $\psi_h$ can be approximately calculated from $\psi_h \approx \psi_{p0} - \psi_{t0}/2 - \alpha_{IS}g$ to represent the perturbed flux surfaces near q=2 surface, where the $\psi_{p0}$ and $\psi_{t0}$ are the unperturbed poloidal and toroidal flux functions, and $g$ is the poloidal current in the covariant Boozer representation ambient magnetic field $\mathbf{B}_0 = \delta \nabla \psi_0 + I \nabla \theta + g \nabla \zeta$.

In this work, the simulations are conducted in the plasma frame, assuming a time scale shorter than the island evolution time. Consequently, we consider the perturbed (2,1) magnetic field $\delta \mathbf{B}_{IS}$ as part of the equilibrium field and focus on the electrostatic turbulence and transport. The particle dynamics are described by the gyrokinetic equation with the parallel symplectic representation of the modern gyrokinetic model. The distribution function $f_s$ of species 's' (ions or electrons) follows the collisionless gyrokinetic Vlasov equation,

$$(L_0 + L_{\delta B} + \delta L)f_s = 0, \tag{1}$$

where

$$L_0 = \frac{\partial}{\partial t} - \frac{\mu}{m_s}\frac{\mathbf{B}_0^*}{B_{0\parallel}^*}\cdot \nabla B_0 \frac{\partial}{\partial v_\parallel} + v_\parallel \frac{\mathbf{B}_0^*}{B_{0\parallel}^*}\cdot \nabla$$
$$+ \frac{\mu}{Z_s B_{0\parallel}^*}\mathbf{b}_0 \times \nabla B_0 \cdot \nabla, \tag{2}$$

$$L_{\delta B} = \left(\frac{B_{0\parallel}^*}{B_\parallel^*} - 1\right)L_0 + v_\parallel \frac{\delta \mathbf{B}_{IS}}{B_\parallel^*}\cdot \nabla - \frac{\mu}{m_s}\frac{\delta \mathbf{B}_{IS}}{B_\parallel^*}\cdot \nabla B \frac{\partial}{\partial v_\parallel}, \tag{3}$$

$$\delta L = \frac{(\mathbf{b}_0 + \delta \mathbf{B}_{IS}/B_0) \times \nabla \delta \bar{\phi}}{B_\parallel^*}\cdot \nabla - \frac{1}{m_s}\frac{\mathbf{B}_0 + \delta \mathbf{B}_{IS}}{B_\parallel^*}\cdot Z_s \nabla \delta \bar{\phi}\frac{\partial}{\partial v_\parallel} - v_\parallel \frac{\nabla \times (\mathbf{b}_0 + \delta \mathbf{B}_{IS}/B_0)\cdot \nabla \delta \bar{\phi}}{B_\parallel^*}\frac{\partial}{\partial v_\parallel}, \tag{4}$$

Where $\mathbf{b}_0 = \mathbf{B}_0/B_0$, $\mathbf{B}_0^* = \mathbf{B}_0 + m_s v_\parallel \nabla \times \mathbf{b}_0/Z_s$, $B_{0\parallel}^* = \mathbf{B}_0^* \cdot \mathbf{b}_0$, $B_\parallel^* = B + (m_s v_\parallel \nabla \times \mathbf{b}/Z_s)\cdot \mathbf{b}$. $m_s$ and $Z_s$ denote the particle mass and charge. $\delta \bar{\phi}$ is the gyroaveraged potential for ions, and we ignore the finite Larmor radius effect for electrons due to small electron gyroradius (i.e., by using electron drift kinetic equation). The $\delta f$ method is used in this work to reduce the numerical noise. The equilibrium distribution is defined by $L_0 F_{s0} = 0$, where $F_{s0}$ is the neoclassical solution in the absence of the MI. And $\delta f_s$ is solved from

$$(L_0 + L_{\delta B} + \delta L)\delta f_s = -(L_{\delta B} + \delta L)F_{s0}, \tag{5}$$

where we use the local Maxwellian to approximate $F_{s0}$ in the $\delta f_s$ equation. The gyrokinetic Poisson equation is used to close the system in the electrostatic simulations.[69], [70]

$$\frac{Z_i^2 n_{0i}}{T_{0i}}(\delta \phi - \delta \tilde{\phi}) = Z_i \delta \bar{n}_i + Z_e \delta n_e, \tag{6}$$

Where $\delta \bar{n}_i$ and $\delta n_e$ are the gyroaveraged guiding center density of ions and electrons. The detailed derivation of the simulation model can be seen in[53]. In simulations without magnetic islands (MIs), it is common practice to solve the electron adiabatic and non-adiabatic responses separately from drift kinetic equation using an iterative method[71], [72]. However, when MIs are present, the situation becomes more complex because electrons move along the perturbed magnetic field lines. In this case, distinguishing between adiabatic and non-adiabatic responses requires separating the electrostatic potential perturbation ($\delta \phi$) into 'zonal' and 'non-zonal' components. This separation is based on whether the local parallel wavenumber ($k_\parallel$) is zero, meaning the parallel derivative along the perturbed magnetic field lines is zero. This process is numerically challenging, as it involves complex field-line-following calculations. To avoid these complications, we solve $\delta \phi$ as a whole, without splitting it into zonal and non-zonal components. As a result, the electron distribution function $\delta f_e$ must also be solved in its entirety using the complete drift-kinetic equation, rather than separating the adiabatic and non-adiabatic components.

In the simulations, we use $260 \times 1200 \times 29$ grid points in the radial, poloidal and parallel direction to simulate the region between $\psi = 0.012\psi_w$ and $\psi = 0.95\,\psi_w$, where $\psi_w$ is the flux function for last closed flux surface. For each grid cell, 100 ions and electrons are loaded initially. The modes of $n \leq 1$ and $9 \leq n \leq 120$ are kept in the simulations. The time step is $\Delta t = 5.0 \times 10^{-4}\,R_0/C_s$, where $R_0$ is the major radius of magnetic axis, and $C_s = \sqrt{T_{e,a}/m_i}$ is the ion acoustic speed on magnetic axis.

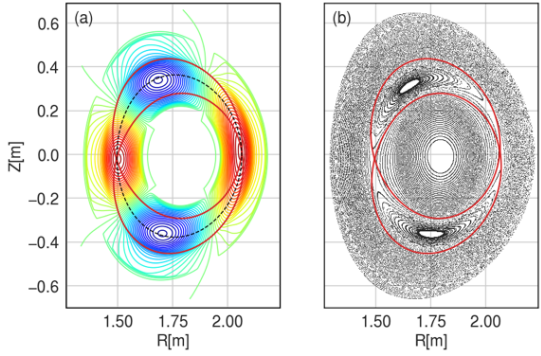

Figure 2 (a) The perturbed magnetic field potential $\alpha_{IS}$ at $\zeta = \pi$ poloidal plane. (b) the Poincare plot of electron guiding center orbits at $\zeta = \pi$ poloidal plane. The red solid lines are the constant $\psi_h$ contour that shows the island separatrix. The dashed line stands for q=2 surface. The same notations are used for the other figures as well.

# 3 simulation results

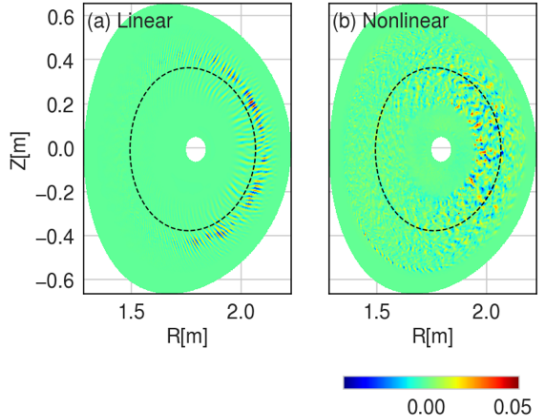

Figure 3 Electrostatic potential $e\phi/T_e$ on poloidal plane in the (a)linear and (b)nonlinear stage of the ITG instability without magnetic island. The color scale for nonlinear potential is shown under panel(b).

To identify the dominant driftwave instability and to delineate the effect of MI, we first exclude the MI field $\delta \mathbf{B}_{IS}$ from the self-consistent turbulence simulation and identify the ion temperature gradient (ITG) mode as the dominant instability. The most unstable eigenmode has n=75 and m$\approx$ 165, $k_\theta \rho_i \sim 0.55$ with the largest amplitude located at the q=2.2 surface (See Fig.3).The linear growth rate and frequency are $\gamma = 6.6 \times 10^4\,s^{-1}$ and $\omega = 3.9 \times 10^4\,s^{-1}$, respectively. In the nonlinear stage, small isotropic eddies form due to the shearing effects of zonal flows. The turbulence fluctuation spreads inward and covers the $q = 2$ region. The averaged heat conductivity between $\rho_t = 0.37$ and $\rho_t = 0.71$ of ions and electrons in nonlinear stage are $\chi_i \approx 2.8\,m^2/s, \chi_e \approx 2.1 m^2/s$, respectively.

We perform the simulation with MI with two stages, the Monte-Carlo stage for pressure relaxation (i.e., only n=1 mode) due to the static MI, and the self-consistent turbulence simulation (i.e., all n modes). Note that the initial density profile in Fig. 1 is not flattened within the island region. If we start the simulation with both MI profile relaxation and self-consistent turbulence, the long time-scale pressure flattening effect emerges due to the equilibrium term $-v_\parallel \frac{\delta \mathbf{B}_{IS}}{B_\parallel^*} \cdot \nabla F_{s0}$ in the RHS of Equation (5). Because of the different flattening of ions and electrons, the non-oscillating charge difference will be created, and consequently an unphysically large electric field is artificially created. This problem is solved by first finding the appropriate initial value of $\delta f$ using Monte-Carlo simulation.

### 3.1 Monte-Carlo simulation

In the Monte-Carlo simulation, $\delta \phi$ is forced to be 0. We solve the perturbed ion distribution function from
$$(L_0 + L_{\delta B})\delta f_{iMC} = -L_{\delta B} F_{i0}, \tag{7}$$
and ions are redistributed spatially under the effect of $\delta \mathbf{B}_{IS}$. Fig. 4(a) shows the time evolution of the entropy, defined as $S_i = \int \delta f_i^2 d^3\mathbf{v} d^3\mathbf{x}$, which indicates the deviation from the initial equilibrium $F_{i0}$. The Monte-Carlo ion simulation is stopped after or 0.4 ms when the entropy reaches its maximum value. The ion profile at the end of the simulation is shown in Fig. 4. In the Monte-Carlo stage (from -0.4 ms to 0 ms), the $S_i$ increases due to flattening effect and eventually reaches a steady value. And in the turbulence stage, $S_i$ continues to increase with the perturbed electric field. The turbulence simulation ends at $t = 0.2$ ms, before the emergence of numerical instability. The density and temperature flattening near q=2 surface can be observed in Fig.4(c) and Fig.4(d). Here, the profile can be calculated using the $\delta f$ or full-$f$ approach and they agree with each other. Then we take $\delta f_{iMC}$ as the initial value of both delta $\delta f_i$ and $\delta f_e$ and start the turbulence simulation by solving Equation (5). Since the $-L_{\delta B} F_{i0}$ term will be cancelled with the initial value of LHS in (5), there is no long time-scale evolution of $\delta f_s$ due to MI and the quasi-neutrality condition is satisfied.

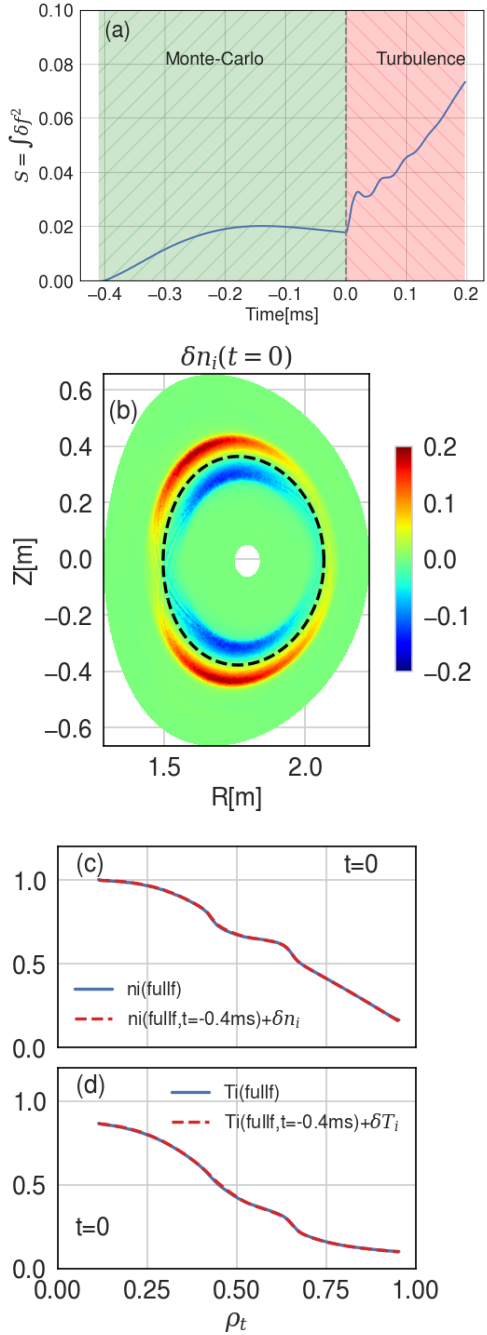

Figure 4 (a) The time history of particle entropy. The Monte-Carlo stage is from t=-0.4ms to t=0, and the turbulence stage is from t=0 to t=0.2ms. (b)(c)(d) The flattened profiles at t=0, after the finish of Monte-Carlo stage. (b) $\delta n_i$ on the $\zeta = \pi$ poloidal plane. (c)Radial ion density profile v.s. $\rho_t$, normalized by $n_e$ on axis. (d)Radial ion temperature profile v.s. $\rho_t$, normalized by $T_e$ on axis. In (c) and (d) the two curves coincide with each other.

## 3.2 Turbulence simulation with MI

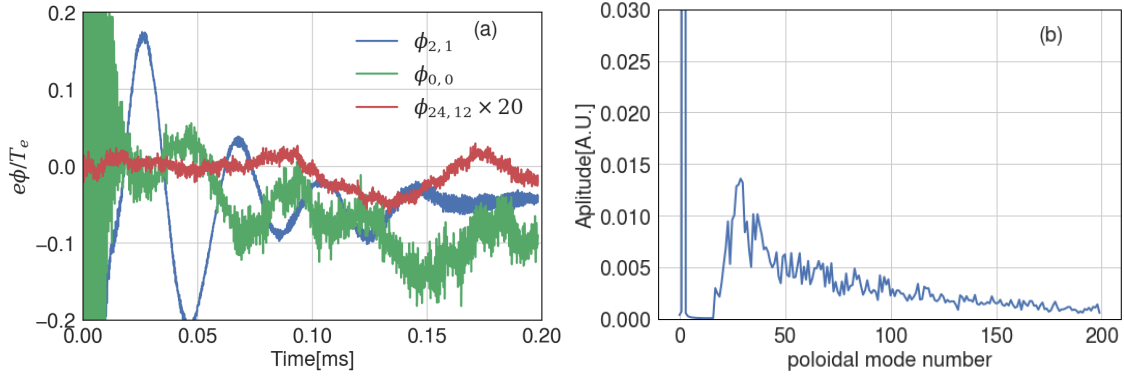

*Figure 5 (a) Time history of ϕ in the simulation with MI at q=2 surface. $\phi_{0,0}$ is the zonal component. $\phi_{2,1}$ is the vortex flow component. $\phi_{24,12}$ is the dominant turbulence fluctuation at q=2 surface. $\phi_{24,12}$ has been amplified by 20 times in the plot for visibility. (b) Relative amplitude of harmonics with different poloidal number at t=0.17ms.*

Fig.5 shows the evolution of electrostatic potential $\phi$, and the relative amplitude of different poloidal mode number at t=0.17ms. Due to the large perturbation from the Monte-Carlo simulation, a large particle noise in $\phi_{00}$ and $\phi_{2,1}$ is created at the beginning of turbulence simulation (but decreases rapidly), and simulation soon enters the nonlinear stage. The evolution of $\phi$ is dominated by the zonal component $\phi_{00}$ and the vortex flow $\phi_{2,1}$. These two components are linearly coupled through the (2,1) magnetic field of the MI, and oscillate at a frequency $\omega = 1.7\, C_s/R_0$, which is slightly lower than the theoretical GAM frequency $\omega_{GAM} \approx 2.1\, C_s/R_0$. From Fig. 5(b), we can see the dominant harmonic of turbulence at q=2 surface is $m = 24, n = 12$, and the time history of $\phi_{24,12}$ in Fig.5(a) shows a much slower oscillation than ITG modes in the case without MI and a much lower amplitude compared to the vortex flow. Further analysis shows the oscillations of the real and imaginary parts of $\phi_{2,1}$ are in phase, which means that the vortex mode is a standing wave, with antinodes located at the X-points and O-points of the islands.

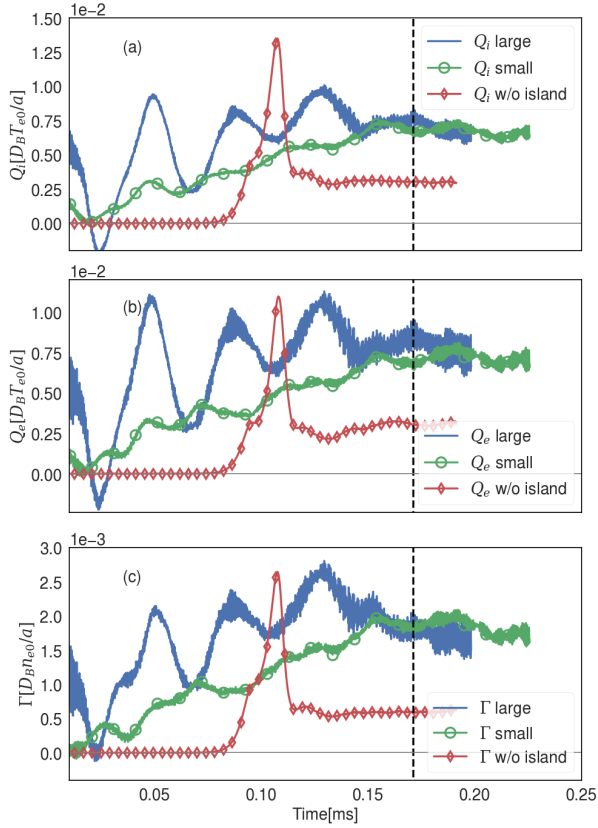

Figure 6 *The time history of Ion heat flux $Q_i$ (panel a), electron heat flux $Q_e$ (panel b), and particle flux $\Gamma$ (panel c).* Here we use $\Gamma \equiv (\Gamma_i + \Gamma_e)/2$ to show the averaging ion and electron particle flux, which should be ambipolar but with some small noises. The vertical dashed line in the three panels corresponds to t=0.17ms, when the snapshots of the turbulence are taken for latter analysis. $Q$ and $\Gamma$ are normalized by $D_B T_{e0}/a$ and $D_B n_{e0}/a$, respectively. $D_B = T_{e0}/eB_0$ is the Bohm unit. $n_{e0}$ and $T_{e0}$ stand for the electron density and temperature on axis. $a$ is the minor radius at plasma wall.

The time history of volume averaged particle and heat fluxes is plotted in Fig. 6. The particle flux is calculated by $\Gamma_s = \int d^3v \, v_r \delta f_s$, and the heat flux is calculated by $Q_s = \int d^3\mathbf{v} \left(\frac{1}{2}mv^2 - \frac{3}{2}T_s\right) v_r \delta f_s / n_{0s}$, where $v_r$ is the radial $E \times B$ velocity, and $n_{0s}$ is the equilibrium density for species 's'. In Fig. 6 the volume average is taken between $\rho_t = 0.35$ and $\rho_t = 0.72$ for all cases. An additional simulation with $\alpha'_{IS} = \alpha_{IS}/2$ was conducted to examine the effect of island size on transport. We refer to the simulation with standard $\alpha_{IS}$ as the 'large island case' and the one with $\alpha'_{IS}$ as the 'small island case'. The transport is clearly modulated by the vortex mode frequency, which is close to the GAM frequency. However, the oscillation is less significant in the small island case due to the weaker vortex mode and smaller island size. In the simulations without island, $Q_i$ and $Q_e$ show a clear linear growth and nonlinear saturation stage. In contrast, in the simulations with islands, the transport level continuously oscillates and increases without a linear growth phase. When the GAM and vortex mode oscillations damp to the residual level, the transport also reaches a steady value. The MI enhances $Q_i$ and $Q_e$ by a factor of 3. Although the oscillation amplitude and the rate of increase of heat transport over time are smaller in the small island case compared to the large island case, the final steady transport level is similar for both cases. The temporal evolution of particle flux $\Gamma$ in Fig.6(c) is similar to heat flux. But importantly, it shows the enhancement of $\Gamma$ by MI, which is consistent with the density pump-out result in RMP experiments. We note that the two simulations here with different island sizes do not systematically address the island size effect using the realistic geometry of the KSTAR tokamak. A systematic study on island size effect on zonal flows and turbulent transport using a simple tokamak geometry can be found in Ref[59], in which a critical island size for turbulence transport enhancement is found. In the future study we will find KSTAR experiments with more comprehensive measurements of the island size and compare the transport with difference island sizes to the GTC simulations.

## 3.3 Turbulence regulation by zonal flows and vortex flows

The instantaneous Er shearing rate of zonal flows can be calculated as $\omega_s = \partial^2 \phi_{00}/\partial \psi^2 \times R_0^2 B_\theta^2/B_0$, which assumes turbulence eddies are isotropic in radial and poloidal direction. The $E_r$ shearing rate averaged over unperturbed flux surfaces from simulations with and without MI are presented in Fig.7(a) and Fig.7(b) as functions of time and radial location. The zonal flow shearing rate has barriers at the island separatrix, particularly at the outer one. Within the island region ($0.42 < \rho_t < 0.65$), $\omega_s$ shows minimal radial dependence and it exhibits damping while oscillating at the GAM frequency. The damping is also observed in the $\phi_{2,1}$ vortex mode in Fig.5. The zonal flow shearing rate within the island region eventually reaches a residual level $|\omega|_s \approx 1.6 \times 10^5 \ s^{-1}$ which is significantly lower compared to the no-island case, where $|\omega|_{s,nois} \approx 1.5 \times 10^6 \ s^{-1}$. However, it is important to note that the effective shearing rate in the no-island case can be reduced due to the finite oscillation frequency of $\omega_s$ [73]. The relation between zonal flow shearing rate and transport can be seen in Fig. 8. The transport is lower where the shearing rate is large in both cases. And the MI has an enhancement for particle and heat flux at all radial locations. From Fig. 7(a) or Fig. 8(b) we see some asymmetry of zonal flow shear at inner and outer separatrices, which may result from different turbulence drive and zonal flow damping at these two locations. It is worth noting that the radial asymmetry for flow shear is also found in some experiments. [16], [19]

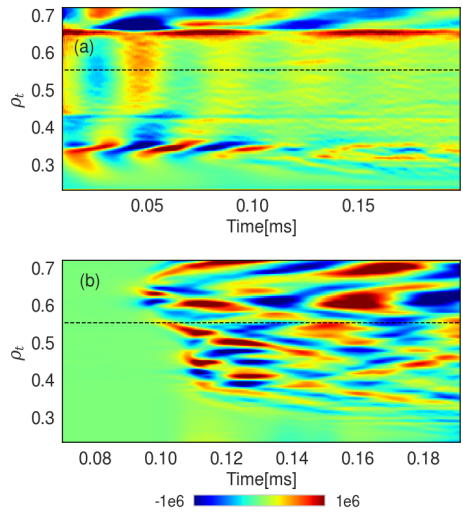

*Figure 7 Zonal flow shearing rate $\omega_s$ as function of simulation time and radial location. The unit of $\omega_s$ is 1/s. (a) With MI. (b) Without MI. The horizontal dashed line shows the q=2 surface.*

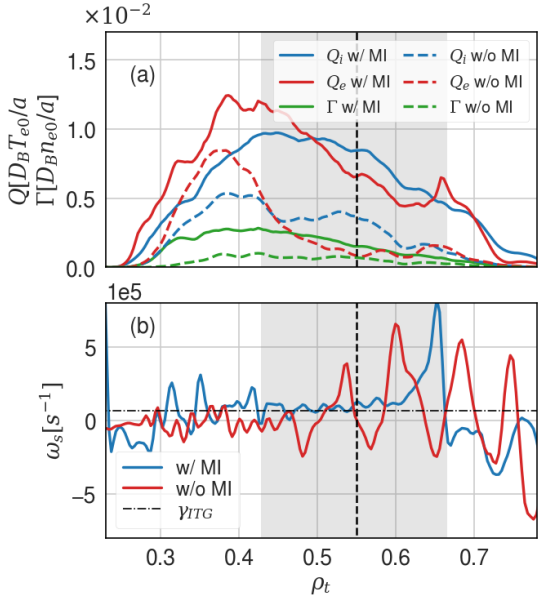

*Figure 8 Time averaged radial profile of particle flux, heat flux, and zonal flow shearing rate $\omega_s$. Time window is [0.16,0.20]ms for case with MI, and [0.14,0.17]ms for case without MI. Q is normalized by $D_B T_{e0}/a$. The linear growth rate of ITG without island ($\gamma = 6.6 \times 10^4\ s^{-1}$) is shown as the horizontal black dash line in panel (b). The gray shaded area shows the island region near O-points. The vertical dashed line shows the q=2 surface.*

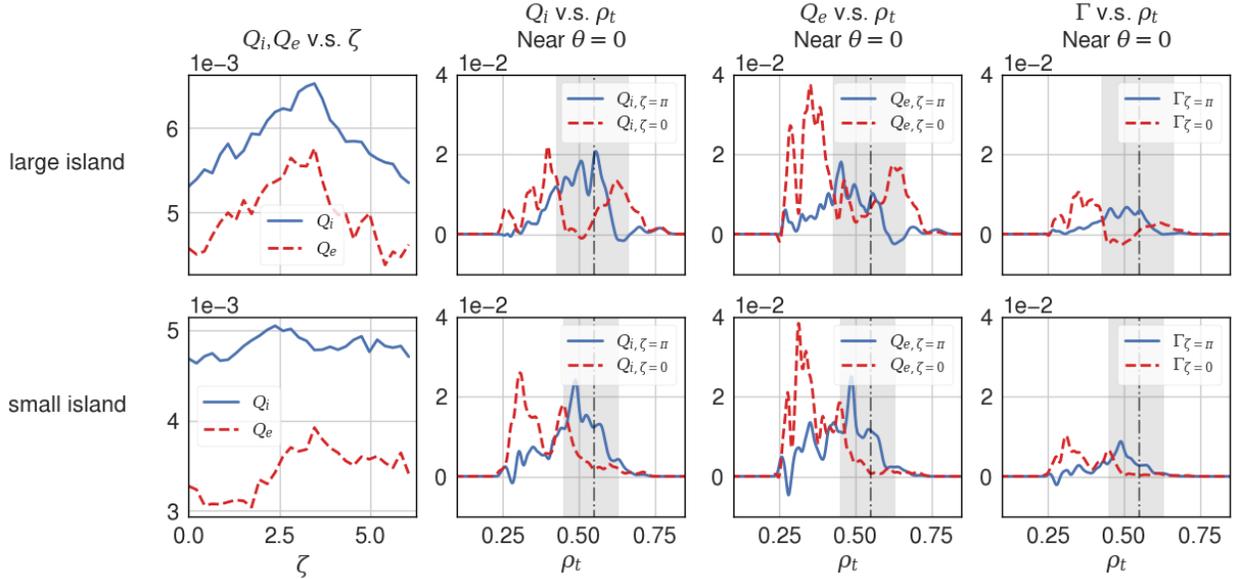

*Figure 9 $Q_i$, $Q_e$, and $\Gamma = (\Gamma_i + \Gamma_e)/2$ when t=0.171 ms (labeled time step in Fig.6). the two rows correspond to the large island case and the small island case, respectively. (First column) Poloidal plane averaged $Q_i$ and $Q_e$ at different $\zeta$ angle, the radial averaging is taken between $\rho_t = 0.42$ and $\rho_t = 0.65$ for large island case, and between $\rho_t = 0.45$ and $\rho_t = 0.62$ for small island case. $\Gamma$ has a similar toroidal variance as Q and not shown in the first column. (Second, third, and fourth column) $Q_i$, $Q_e$, and $\Gamma$ averaged between $\theta \in [-\pi/8, \pi/8]$, when $\zeta = 0$ (Near O-point) or $\zeta = \pi$ (Near X-point). Shown as a function of $\rho_t$. Shaded area shows the island region near O-points, and the vertical dashed line shows q=2 surface. Q and $\Gamma$ are normalized by $D_B T_{e0}/a$ and $D_B n_{0e}/a$, respectively.*

To further examine effect of MI on the transport spatial structures, we plot the radial and toroidal variance of particle and heat transport in Fig.9. When t=0.171 ms, the vortex flow damps to a residual level and the transport reaches a steady level. In the first row, the first column indicates that $Q_i$ and $Q_e$ in the large island case exhibit a clear toroidal angle dependence. The radial dependence of $Q$ and $\Gamma$ near outer mid plane ($\theta = 0$) are plotted in the next three columns, where we see the clear transport concentration at X-point, and the transport suppression at O-point. We should note that the toroidal dependence is not only due to the difference between O-point and X-point, but also due to the difference of X-points at $\zeta = 0$ and $\zeta = \pi$. In the small island case (second row of Fig.9), the toroidal variance of transport is less clear. Although the transport is suppressed in the island region near O-point, the strong transport level is found at smaller $\rho_t$. Besides, the transport level near the X-points at $\zeta = 0$ and $\zeta = \pi$ is similar. Thus, the differences between poloidal planes at different $\zeta$ angles are reduced.

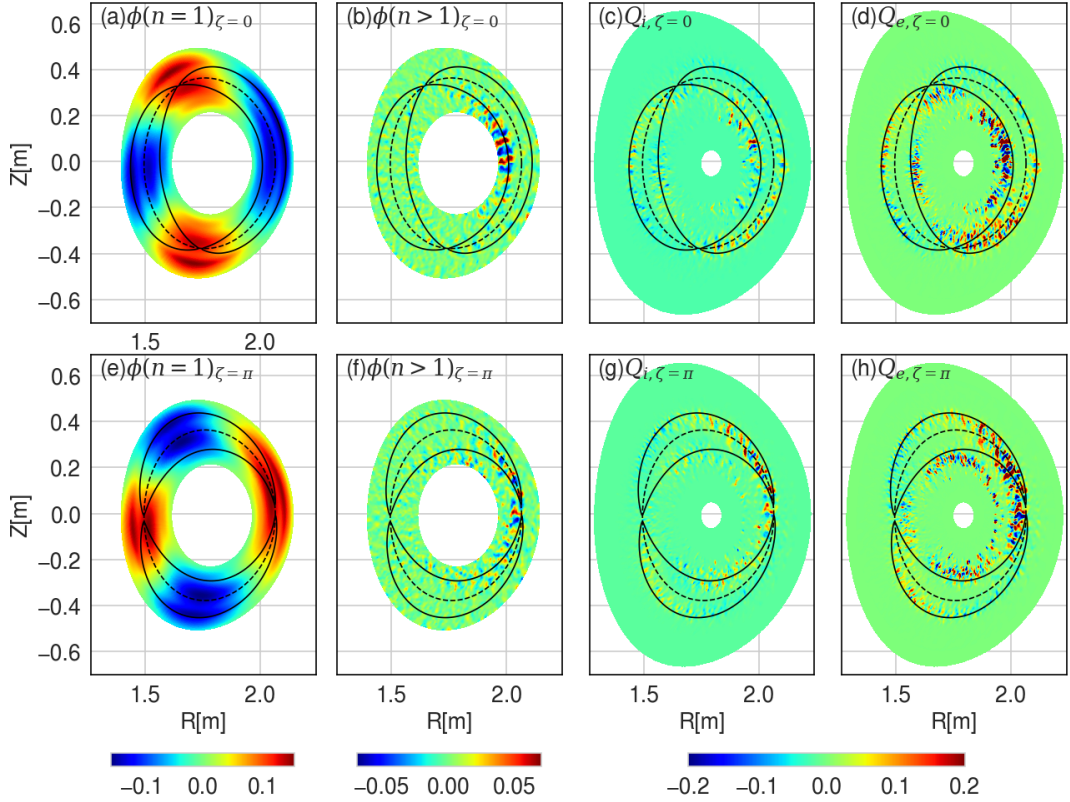

*Figure 10 The electrostatic potential and heat flux at t=0.17ms. (a) and (e) $\phi(n = 1)$ component at $\zeta = 0$ and $\zeta = \pi$. (b) and (f) turbulence component of $\phi$ at $\zeta = 0$ and $\zeta = \pi$, $\phi(n > 1) = \phi - \phi_{00} - \phi(n = 1)$. (c) and (g) ion heat flux $Q_i$ at $\zeta = 0$ and $\zeta = \pi$. (d) and (h) electron heat flux at $\zeta = 0$ and $\zeta = \pi$. The potential is normalized by $T_{e0}/e$, and the $Q$ is normalized by $D_B T_{e0}/a$.*

Interplays between vortex flows and turbulence are illustrated in Fig. 10, which shows the snapshots of vortex flow $\phi(n = 1)$ and turbulence component of potential $\phi(n > 1)$ and radial heat flux $Q_i$ and $Q_e$ during the turbulence simulation when $t = 0.17$ ms. $\phi(n = 1)$ simply rotates along the field line and structure does not depend on $\zeta$ angle, while the different effect of MI at different toroidal angles is observed for turbulence fluctuation and associated transport. At $\zeta = 0$, the O-point is located at mid-plane, and the potential fluctuation is maximum near the inner separatrix at the outer mid-plane ($\theta = 0$). The heat flux is mainly along the separatrix, especially at the region near X-points. This observation shows the effect from the vortex flow shear, which is qualitatively the same as the results in the analytical geometry[53], [54] . Compared to the analytical geometry, the simulations in KSTAR geometry show smaller eddy size, and at $\zeta = 0$ poloidal plane, the turbulence fluctuation of $\phi$ near X-points is not strong since it is not in the bad curvature region. At $\zeta = \pi$, the maximum values of potential fluctuation and the heat flux are found near the X-point at $\theta = 0$. Compared with the poloidal structure of $\phi$ at earlier time points (not shown here) in the simulation, the turbulence spreading into the island region through X-point can be clearly observed near the X-point at $\theta = 0$. Compared to the $\zeta = 0$ plane, the turbulence amplitude is larger near this X-point, since the ITG turbulence drive is stronger at outer mid-plane

(see Fig.3 (b) ). Consequently, the turbulent transport across q=2 surface is also stronger at $\zeta = \pi$ than $\zeta = 0$. This observation is consistent with the 1-dimensional analysis in Fig.9.

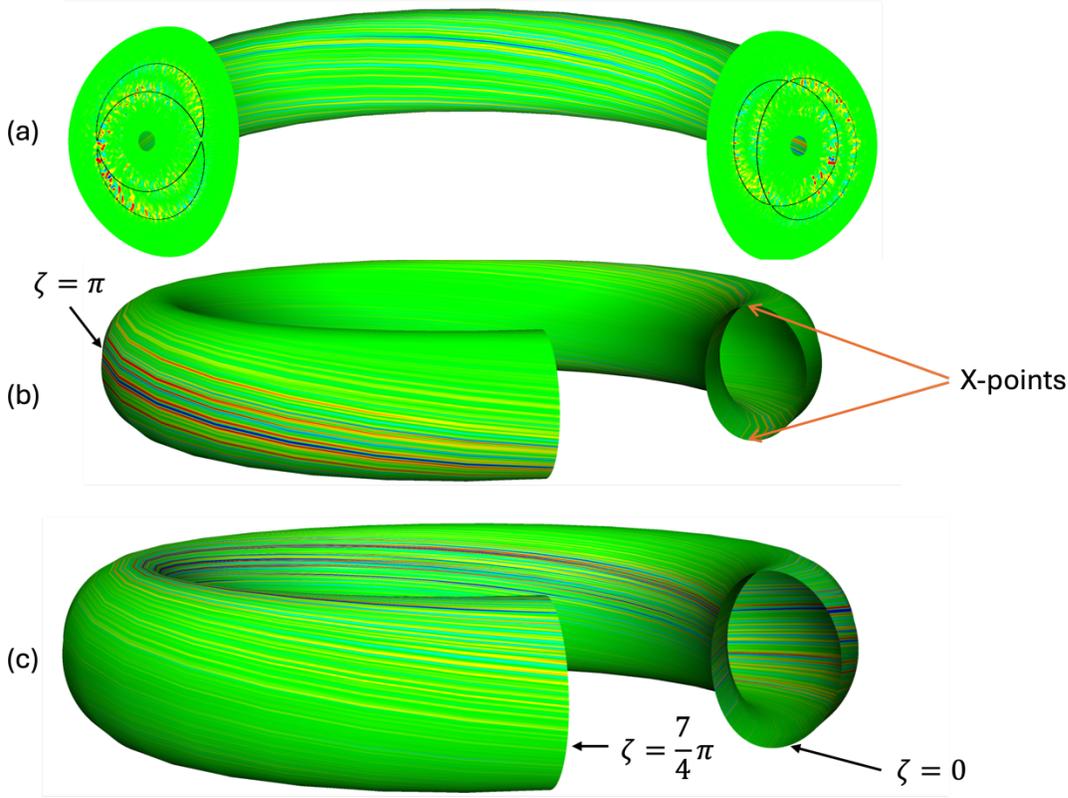

Figure 11 $Q_i$ on different flux surfaces when t=0.17ms. (a) $\rho_t = 0.42$(near inner separatrix) (b) $\rho_t = 0.55$ (q=2 surface) (c) $\rho_t = 0.65$ (near outer separatrix). $Q_i$ is normalized by $D_B T_{e0}/a$. The poloidal planes at $\zeta = 0$ and $\zeta = \pi$ are shown in panel(a). The torus from $\zeta = 0$ to $\zeta = 7/4\,\pi$ is shown in (b) and (c). The two X-points at $\zeta = 0$ are labeled in panel(b).

The vortex flow effect on the transport can also be clearly seen from a 3-D plot in Fig. 11. Three flux surfaces are selected with q=1.69, q=2, and q=2.33, and $Q_i$ are plotted on these surfaces. Here, the $q = 1.69$ and $q = 2.33$ surfaces are selected to be tangent to the inner and outer separatrix, respectively. In Fig.11(a) and Fig.11(c), the $\theta$ value corresponding to the maximum $Q_i$ at $q = 1.69$ and $q = 2.33$ surfaces varies toroidally, matching the phase of island, but the poloidally averaged amplitude of $Q_i$ does not depend on the toroidal angle significantly. In Fig. 11(b), the maximum region for $Q_i$ at $q = 2$ appears only near $\zeta = \pi$ and $\theta = 0$, which is in consistency with the toroidal varying $Q$ in Fig. 10. This again shows the transport is easier to cross the $q = 2$ surface near $\zeta = \pi$. Thus, the toroidal variance of the fluctuation and transport property is created in a tokamak with MIs.

## 3.4 Comparison with experiments

The ion and electron density fluctuations at the $\zeta = \pi$ poloidal plane are shown in Fig.12. The similarity in densities between the two species demonstrates the maintenance of quasi-neutrality condition during the simulation. We measured the density fluctuation level at q=2 surface, defined by $\delta \tilde{n} = \delta n - \delta n_{00} - \delta n_{2,1}$, and found the fluctuation is suppressed at the O-points. This fluctuation regulation effect from the MI is consistent with previous observations in TEXT[67] and DIII-D[66] experiments. Additionally, we show the two X-points are not equivalent, and the fluctuation is only maximum when $\theta = 0$.

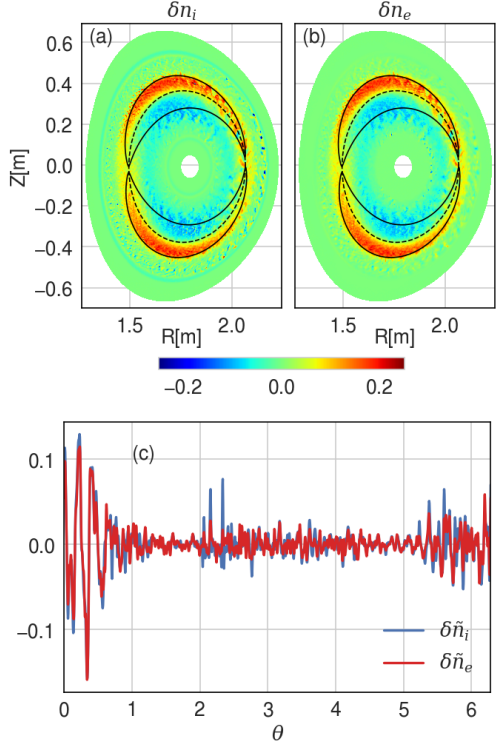

*Figure 12 The density fluctuations on the $\zeta = \pi$ poloidal plane of ions(a) and electrons(b) when t=0.17ms. (c) The density fluctuations at q=2 surface dependence on poloidal angle. Density is normalized to the electron density on magnetic axis.*

Fig.13 compares the frequency spectrum of the turbulence between KSTAR experiments and GTC simulations. In the KSTAR discharge #19115 with RMP suppressing ELM, the turbulence fluctuation frequency near the X-point was measured using Microwave Imaging Reflectometry (MIR)[74], [75]. The experimental conditions are similar to those in the simulated discharge #19118[64], [65]. In the KSTAR experiments, the MIR channels are placed at outer mid-plane near R=2 m. In an experiment with rotating RMP, the X-point and O-point will be seen by MIR sequentially at $\theta = 0$. In this simulated case, the RMP is not rotating. If we define the toroidal angle of X-point with $\theta = 0$ as $\zeta = \pi$, then the MIR and ECEI measurements are located at $\zeta = 9\pi/8$ and $\zeta = 7\pi/8$, respectively. The X-points that have the same toroidal location with ECEI are located at $\theta \approx -\pi/4$ and $\theta \approx 3\pi/4$ (See Fig.1(b)). The X-points that have the same toroidal location with MIR are located at $\theta \approx \pi/4$ and $\theta \approx 5\pi/4$. The MIR measurements were recorded at two time points: t = 2.8 s (before RMP and mode locking), labeled as 'MIR w/o island', and t=3.05s (after RMP, with mode locking) labeled as 'MIR w/ island'. The corresponding frequency spectrum from GTC simulations is obtained by conducting the Fourier transformation to the dominant turbulence harmonics. In Fig.13, the blue line represents the GTC simulation with the island, and the red line represents the simulation without the island. Prior to the RMP (orange line), a peak frequency at 150 kHz corresponds to the toroidal rotation frequency. After the formation of the magnetic island (MI), the mode becomes locked, and the peak frequency drops to 0. GTC simulations are performed in a rotating frame, the red line in Fig.13 is horizontally shifted by the toroidal rotation frequency for better visibility. A reduction of the width of the spectrum is observed in both experiments and simulations. The standard deviation of the frequency spectrum, which quantifies the width, was calculated and shows consistency between the experiments and simulations, as listed in Table 1. Note that GTC simulations exhibit a high-frequency tail, especially for the case with island, which is likely caused by numerical noise and the finite simulation time. Therefore, the frequency range was restricted in the calculation of the standard deviation: from -100 kHz to 100 kHz for the "with island" case, and from -200 kHz to 500 kHz for the "no island" case. It is worth noting that the MI effect on both the spectrum shape and the width agrees quantitatively between simulations and experiments, which is reported for the first time.

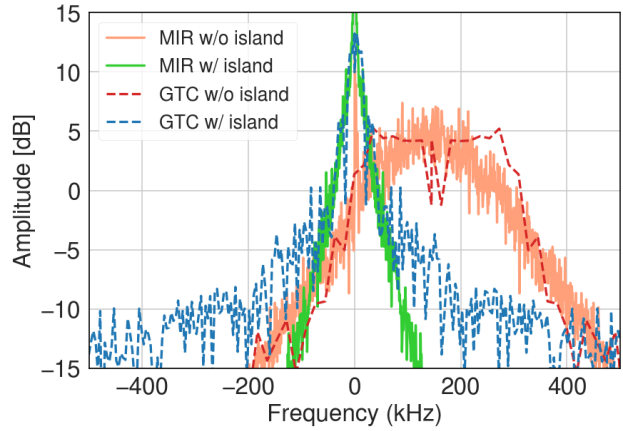

*Figure 13 Comparison of turbulence frequency spectrum between MIR measurement of KSTAR discharge #19115 and GTC simulation. Dashed lines are from simulation (labeled as 'GTC') and solid lines are from MIR data (labeled as 'MIR'). The red line has been shifted horizontally to match the rotation frequency of the orange line (rotation frequency). Both the blue line and red line have been shifted vertically to match the peaked amplitude with the MIR measurement.*

*Table 1 Comparison of the spectrum width in Fig.13.*

|  | With MI | Without MI |
|---|---|---|
| MIR spectrum width (kHz) | 99.6 | 19.2 |
| GTC spectrum width (kHz) | 99.3 | 25.2 |

In addition, we have compared the perpendicular wavenumber $k_\perp$ spectrum between GTC simulation of #19118 and MIR measurement of #19115 in Fig. 14. The results from simulation and experiment agree well. In the case without MI, the turbulence is driven by ITG instability and the $k_\perp \rho_i$ spectrum goes through a reverse cascading process in the nonlinear stage and eventually peaked at $k_\perp \rho_i \approx 0.25$. While in the case with MI, the amplitude of the vortex mode is larger than the ITG turbulence, and the $k_\perp$ is dominated by the $k_r$ of the vortex mode, which can be estimated from the island width, $k_r \rho_i \sim \rho_i 2\pi/w_{is} = 0.11$. Thus the $k_\perp \rho_i$ spectrum with island is much narrower. The harmonics with $n \in [2, 9]$ in the simulations are not included for better numerical properties. This operation can be regarded as a numerical damping around $k_\perp \rho_i \sim 0.15$ and makes the peak for the blue dashed line in Fig. 14 around $k_\perp \rho_i \approx 0.2$. Note that the dominance of vortex mode does not directly introduce the transport enhancement. The spatial structure of $Q_i$ and $Q_e$ in Fig. 10 indicates that the transport is mainly driven by turbulence, which is regulated by the zonal flow and vortex flow with the presence of MIs.

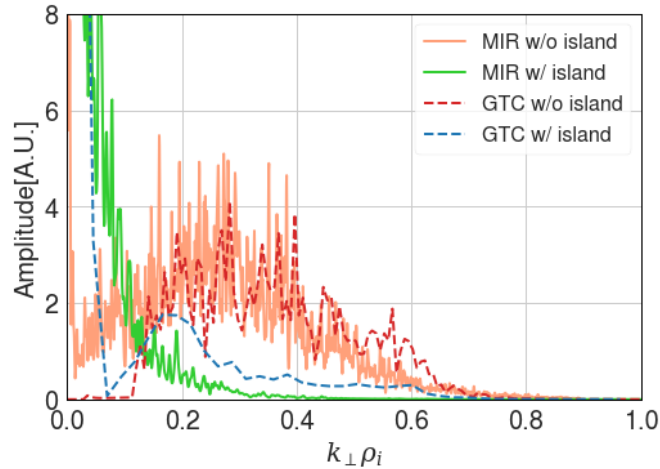

*Figure 14 Comparison of $k_\perp \rho_i$ spectrum between MIR measurement of KSTAR discharge #19118 and GTC simulation. Dashed lines are from simulation (labeled as 'GTC') and solid lines are from MIR data (labeled as 'MIR').*

## 4 Summary

In this study, we have applied the gyrokinetic approach to investigate the effects of MIs on ITG turbulent transport in the KSTAR tokamak. By utilizing a separated Monte-Carlo stage and a self-consistent turbulence simulation stage, we achieved a flattened profile while maintaining quasi-neutrality in the turbulence simulation. We observed the generation of the m=2, n=1 vortex flow from the coupling between GAM and the static MI field. In the presence of MIs, the turbulence potential oscillates at a lower frequency than the ITG frequency in the case without island. MIs enhance the particle and heat transport in both ion and electron channels, which is further modulated by vortex flows and oscillates at the GAM frequency. Zonal flows are predominantly generated outside the island region, with a significant flow shear near the island separatrix. Under the effect of zonal flows and vortex flows, the transport is redistributed spatially. On a poloidal plane, the radial heat transport is concentrated along the island separatrix but also penetrates in the island region via the X-points. Transport along the island separatrix shows toroidal dependence during the vortex flow oscillation. When the vortex flow damps, transport at $q = 2$ surface is more intense at $\zeta = \pi$ than $\zeta = 0$, where one of the X-points is located at the outer-mid plane (bad curvature region). These findings suggest that the confinement can be improved more efficiently by suppressing the transport near $q = 2$ and $\zeta = \pi$. For the first time, the quantitative agreement between simulations and experiments have been shown on the MI impact on frequency and perpendicular wavenumber spectrum. In the current work, we found the heat transport driven by ITG turbulence is much larger than the particle transport within the island region, which is not consistent with the measurement near the edge. Besides, we only include the m=2, n=1 resonant component of $\delta \mathbf{B}$. In the future work, we will compare the impact on transport from resonant and non-resonant components. We will include the plasma boundary region to compare the particle transport driven by turbulence and the density pump-out measurement.

## Acknowledgement

We would like to thank T.S. Hahm for usefusl discussions. This work is supported by US-DOE under grant DE-SC0023434 and DE-FG02-07ER54916, and used resources of the Oak Ridge Leadership Computing Facility at Oak Ridge National Laboratory (DOE Contract No. DE-AC05-00OR22725) through an INCITE award and the National Energy Research Scientific Computing Center (DOE Contract No. DE-AC02-05CH11231). This report was prepared as an account of work sponsored by an agency of the United States Government. Neither the United States Government nor any agency thereof, nor any of their employees, makes any warranty, express or implied, or assumes any legal liability or responsibility for the accuracy, completeness, or usefulness of any information, apparatus, product, or process disclosed, or represents that its use would not infringe privately owned rights. Reference herein to any specific commercial product, process, or service by trade name, trade- mark, manufacturer, or otherwise, does not necessarily constitute or imply its endorsement, recommendation, or favoring by the United States Government or any agency thereof. The views and opinions of authors expressed herein do not necessarily state or reflect those of the United States Government or any agency thereof.